\documentclass{pasj00}
\begin{document}
\SetRunningHead{Author(s) in page-head}{Running Head}
\Received{2006/07/05}
\Accepted{2006/08/16}

\title{Iron and Nickel Line Diagnostics for the Galactic Center Diffuse 
Emission}

\author{Katsuji \textsc{Koyama}, Yoshiaki \textsc{Hyodo}, Tatsuya \textsc{Inui}, Hiroshi \textsc{Nakajima}, \\
Hironori \textsc{Matsumoto} and Takeshi Go \textsc{Tsuru}}
\affil{Department of Physics, Graduate school of Science, Kyoto University, 
Sakyo-ku, Kyoto 606-8502}
\email{koyama@cr.scphys.kyoto-u.ac.jp, hyodo@cr.scphys.kyoto-u.ac.jp}

\author{Tadayuki \textsc{Takahashi}, Yoshitomo   \textsc{Maeda} and Noriko  \textsc{Yamazaki}} 
\affil{Institute of Space and Astronautical Science, JAXA, Sagamihara, Kanagawa, 229-8510}

\author{Hiroshi \textsc{Murakami}}
\affil{PLAIN center, ISAS/JAXA, 3-1-1 Yoshinodai, Sagamihara, Kanagawa 229-8510}

\author{Shigeo \textsc{Yamauchi}}
\affil{Faculty of Humanities and Social Sciences, Iwate University,
3-18-34 Ueda, Morioka, Iwate 020-8550}

\author{Yohko \textsc{Tsuboi}}
\affil{Department of Physics, Faculty of Science and Engineering, Chuo University, 1-13-27 Kasuga, \\
Bunkyo-ku, Tokyo 112-8551, Japan}

\author{Atsushi \textsc{Senda}}
\affil{The Institute of Physical and Chemical Research (RIKEN),
 2-1 Hirosawa, Wako, Saitama 351-0198}

\author{Jun \textsc{Kataoka}}
\affil{Department of Physics, Tokyo Institute of Technology, Meguro, Tokyo, 152-8551}

\author{Hiromitsu \textsc{Takahashi}}
\affil{Department of Physical Science, School of Science, Hiroshima University,
1-3-1 Kagamiyama, \\
Higashi-Hiroshima, Hiroshima 739-8526}

\author{Stephen S \textsc{Holt}}
\affil{Olin College, Needham, MA 02492 USA}

\and

\author{Gregory V \textsc{Brown}}
\affil{High Energy Density Physics and Astrophysics Division, Lawrence Livermore National Laboratory,\\
 Livermore, California 94551, USA}
\maketitle

\begin{abstract}
We have observed the diffuse X-ray emission from the Galactic center (GC) using the X-ray Imaging Spectrometer (XIS) on Suzaku. 
The high-energy resolution and the low-background orbit provide excellent spectra of the GC diffuse X-rays (GCDX). 
The XIS found many emission lines in the GCDX near the energy of K-shell transitions of iron and nickel. 
The most pronounced features are Fe\emissiontype{I}~K$\alpha$ at 6.4~keV  and K-shell absorption edge at 7.1~keV, which are from neutral and/or low ionization states of iron, and the K-shell lines at 6.7~keV and 6.9~keV from He-like (Fe\emissiontype{XXV}~K$\alpha$) and 
hydrogenic (Fe\emissiontype{XXVI}~Ly$\alpha$) ions of iron. 
In addition, ~K$\alpha$ lines from neutral or low ionization nickel (Ni\emissiontype{I}~K$\alpha$) and He-like nickel
(Ni\emissiontype{XXVII}~K$\alpha$), 
and Fe\emissiontype{I}~K$\beta$, Fe\emissiontype{XXV}~K$\beta$,
 Fe\emissiontype{XXVI}~Ly$\beta$,  Fe\emissiontype{XXV}~K$\gamma$
 and Fe\emissiontype{XXVI}~Ly$\gamma$ are detected for the first time. 
The line center energies  and widths of Fe\emissiontype{XXV}~K$\alpha$ and Fe\emissiontype{XXVI}~Ly$\alpha$ favor a collisional excitation (CE) plasma for the origin of the GCDX. 
The electron temperature determined from the line flux ratio of Fe\emissiontype{XXV}-K$\alpha$/ Fe\emissiontype{XXV}-K$\beta$  is similar to the ionization temperature determined from that of 
Fe\emissiontype{XXV}-K$\alpha$/Fe\emissiontype{XXVI}-Ly$\alpha$. 
Thus it would appear that the GCDX plasma is close to ionization equilibrium.  
The 6.7~keV flux and temperature distribution to the galactic longitude is smooth and monotonic, 
in contrast to the integrated point source flux distribution. 
These facts support the hypothesis that the GCDX is truly diffuse emission rather than the integration of the outputs of a large number of unresolved point sources. 
In addition, our results demonstrate that the chemical composition of Fe in the interstellar gas near the GC is constrained to be about 3.5 times solar.
\end{abstract}

\section{Introduction}
The iron K-shell lines near the Galactic center (GCDX) was first detected with the Ginga satellite \citep{Ko89, Ya90}. 
The line energy is about 6.7~keV and the emission extends about 1--2$^\circ$ over the Galactic center (GC) region.
The ASCA satellite observed more details of the GC region, and confirmed the nature of the GCDX \citep{Ko96,Ta00}. 
Moreover, the ASCA CCD resolved  the iron K-shell line detected with Ginga into three distinct lines at
6.4, 6.7 and 6.9~keV \citep{Ko96}. 
The 6.7 and 6.9~keV lines are likely due to He-like (Fe\emissiontype{XXV}) and hydrogenic (Fe\emissiontype{XXVI}) ions of iron. 
There are three plausible scenarios that have been proposed for the origin of the 
iron K-shell lines. One is collisional excitation  (CE)  by low energy electrons (LEE), similar to the Galactic ridge emission. 
This process, however, contributes mainly to the 6.4 keV line emission, and not to the 6.7 and 6.9~keV lines (Valinia et al. 2000).
With respect to the 6.7 keV and 6.9 keV lines, CE in a thin, high temperature plasma, and charge exchange (CX) recombination are both considered.
For the CE case, the best-fit thin thermal plasma temperature of the GCDX spectrum was $\sim$10~keV.
If the GCDX is really due to a thin hot and uniformly distributed plasma, the total thermal energy is estimated to be 10$^{53-54}$~erg, or equivalent to 10$^2$--10$^3$ supernovae explosions. 
The temperature of $\sim$10~keV is higher than that bounded by the Galactic gravity, hence the plasma should escape from the GC region. 
The time scale estimated by the plasma size (1$^\circ$ or $\sim$ 150~pc at 8.5 kpc) divided by the sound velocity of the $\sim$10~keV plasma 
is $\sim$10$^5$ years. Therefore the energy of $\sim$10$^{53-54}$~erg should be supplied in the past $\sim$10$^5$ years, 
e.g. 1 SN every 100--1000 year is required in the GC region. 
This huge energy budget in the GC region suggests that CX recombination may be a more reasonable explanation for the GCDX.  
In this scenario, during a collision between neutral hydrogen and Fe\emissiontype{XXVI} or Fe\emissiontype{XXVII},  the bound hydrogen electron 
is captured into an excited state of the iron, and then the excited state decays producing an X-ray. 
The X-ray spectral signature is distinct from the signature produced by CE  in a  collisional ionization equilibrium (CIE) plasma, such as the high-temperature plasma mentioned above.  
In the CX case, the forbidden line at $E_{\rm f}=6636$~eV, is stronger than the resonance line at $E_{\rm r}=6700$~eV \citep{Ot06,Be05,Wa05}, 
whereas the resonance line is stronger in the case of a plasma in CIE. 
Thus, even if the resonance and forbidden lines are not completely resolved, the energy centroid of the K$\alpha$ line of Fe\emissiontype{XXV} 
produced by CX will be lower than produced by CE.
We also note that an energy  shift towards lower energy as  the CX  would be observed in the case of X-ray emission following recombination with low energy electrons, as in a photo-ionized plasma \citep{Be03}.  

The 6.4~keV line is likely to be K$\alpha$ from the neutral and/or a blend of  low charge states of iron (here, Fe\emissiontype{I}). 
In fact, the 6.4~keV line is localized near molecular clouds such as the radio complex Sgr B2 and the Radio Arc regions. 
Sgr B2, the most massive giant molecular cloud with ultra compact H\emissiontype{II} (UCH\emissiontype{II}) and many maser sources, 
has been extensively studied with ASCA and Chandra (\cite{Ko96};  \cite{Mu00}, 2001). 
As a result of those studies, the authors proposed that the 6.4~keV line arises in an X-ray reflection nebula (XRN) irradiated by Sgr A$^*$ during an exceptionally bright period that occurred about 300 years ago, i.e., a time span equivalent to  the light traveling time between Sgr B2 and Sgr A$^*$. This putative high activity of Sgr A$^*$ could provide the energy supply to the GCDX.

Accordingly, a high quality observation of the iron and nickel K-shell lines can provide key information for understanding
the origin of the GCDX.  Using the X-ray Imaging Spectrometer (XIS) (Koyama et al.~2006a),
we have measured  X-ray emission for the spectral band containing these lines.
The XIS is the X-ray CCD camera system on board Suzaku (\cite{Mi06}). 
The excellent energy resolution and well-calibrated performance for diffuse sources, coupled with the low background orbit of Suzaku,
make the XIS well-suited to provide crucial information about the GCDX. Here we present 
the results of the first measurement of the GCDX with the XIS focusing on the K-shell emission from iron and nickel.

\section{Observations and Data Processing} 
   \begin{table*}[!ht]
          \begin{center}
     \caption{Suzaku on-source observations near the Galactic center.}
       \label{tbl:gc_log}
\begin{tabular}{lllllll}
         \hline\hline
         Target name  &  Seq. No.     & \multicolumn{2}{c}{Pointing direction}
& Observation Start & \multicolumn{2}{c}{Effective Exposure}\\
         &               & $\alpha$(J2000)&$\delta$(J2000)        &   
&   FI(ks)&BI(ks)       \\
         \hline
         GC\_SRC1   &100027010&\timeform{17h46m03s} &$-$\timeform{28D55'32''}
&2005-09-23\ 07:07&48.6&48.3\\
         GC\_SRC2   &100027020&\timeform{17h45m13s} &$-$\timeform{29D10'16''}
&2005-09-24\ 14:16&40.0&46.6\\
         GC\_SRC2   &100037010&\timeform{17h45m13s} &$-$\timeform{29D10'16''}
&2005-09-29\ 04:25&47.6&47.7\\
         GC\_SRC1   &100037040&\timeform{17h46m03s}  &$-$\timeform{28D55'32''}
&2005-09-30\ 07:41&47.1&47.1\\
         \hline \\
       \end{tabular}     
\end{center}
   \end{table*}

Two pointing observations towards the GC were performed with the XIS at the focal planes of the X-Ray Telescopes (XRT) on board the Suzaku satellite in September of 2005. 
Details of Suzaku, the XRT, and the XIS are found in \citet{Mi06}, \citet{Se06} and Koyama et al. (2006a), respectively. 
Data taken at elevation angles less than 5$^\circ$ from the Earth rim or during the passage through the South Atlantic Anomaly were discarded. 
After this filtering, the net exposure times were $\sim$180~ks and $\sim$190~ks for the front-illuminated (FI) CCDs and the back-illuminated (BI) CCD, respectively. The observation logs are listed in table \ref{tbl:gc_log}. 

Since the GCDX extends beyond the field of view (FOV) of the XIS, the strong K-shell lines are found in the full imaging area (IA) of the XIS CCD. 
We therefore made a fine correction of the charge transfer inefficiency (CTI) and fine gain tuning of CCD-to-CCD and segment-to-segment levels (for the CTI and the CCD segment, see Koyama et al. 2006a). 
These fine tunings were done using the K$\alpha$ lines of Fe\emissiontype{XXV} (6.7~keV),  Fe\emissiontype{I} (6.4~keV) and helium-like  silicon (S\emissiontype{XV}) (2.46~keV). 
The absolute gain tuning was made using the $^{55}$Fe calibration sources irradiating the CCD corners. Details of the fine tuning are given in the Appendix.

\section{Analysis and Results} 

\subsection{The Image and Spectrum near the Iron and Nickel K-Line Complexes} 

\begin{figure}[!ht]
   \begin{center}
     \FigureFile(70mm,50mm){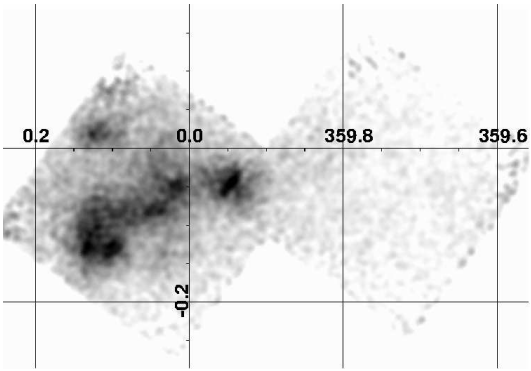} 
     \caption{The narrow band map at the 6.4 keV line (the 6.34--6.46~keV band). Coordinates are 
galactic $l$ and $b$ in degrees.}
     \label{fig:6.4 keV}
\end{center}
\end{figure}

\begin{figure}[!ht]
   \begin{center}
     \FigureFile(70mm,50mm){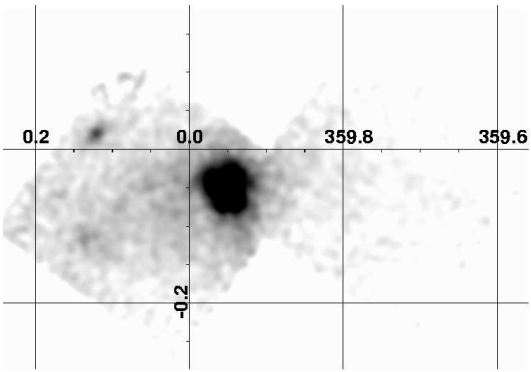} 
     \caption{Same as figure \ref{fig:6.4 keV}, but for the 6.7 keV line (the 6.62--6.74~keV band).}
     \label{fig:6.7 keV}
\end{center}
\end{figure}

After the CTI-correction and the fine gain-tuning, we co-added the four XIS data sets and made X-ray mosaic maps of the GC region. 
Figures \ref{fig:6.4 keV} and \ref{fig:6.7 keV} show the narrow band maps in the 6.34--6.46~keV and 6.62--6.74~keV bands, where the non-X-ray background (NXBG) was
 subtracted and the vignetting correction was applied.
These represent the K$\alpha$ lines of Fe\emissiontype{I} and Fe\emissiontype{XXV}. The sharp contrast in these two adjacent lines demonstrates that the GCDX exhibits a large variety in different emission lines.

\begin{figure}[!ht]
   \begin{center}
\FigureFile(70mm,50mm){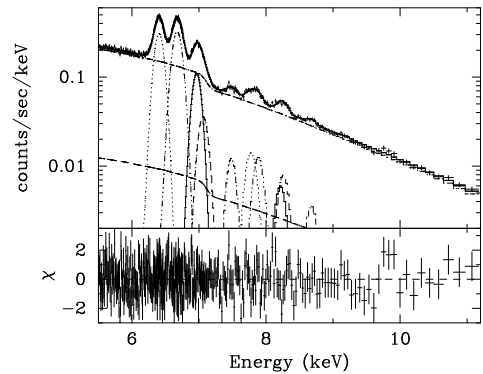} 
     \caption{The averaged spectrum of the 3 FI-CCDs in the 5.5--11.5~keV band. 
The data are taken from the full FOV excluding the calibration source 
regions and Sgr A East. 
The spectra of the three FIs and BI are simultaneously fitted with a model of 
a thermal bremsstrahlung plus ten Gaussian lines and an iron absorption edge. 
The best-fit result of the three-FI spectrum is only shown. 
The long-dashed line is the model of the cosmic X-ray background (CXB).}
     \label{fig:hardspec}
\end{center}
\end{figure}

We made the X-ray spectrum from the whole region of figures \ref{fig:6.4 keV} and \ref{fig:6.7 keV} excluding the corners of the CCD, containing the calibration sources and the central bright object 
Sgr A East (\cite{Ma02, Sa04, Pa05}; Koyama et al. 2006b). 
Since the relative gain of each XIS and segment is well calibrated (see Appendix), and the response functions for each XIS are essentially the same, we added the spectra of the three front-illuminated (FI) CCDs (XIS0, 2 and 3) to increase photon statistics. 
To estimate the NXBG, we used the night Earth data accumulated  by the XIS team (Koyama et al. 2006a).
 The data are sorted with the geomagnetic cut-off rigidity (COR), hence we compiled the data set such that the COR distribution becomes the same as that of the source observation. In reality, the NXBG flux sorted with the COR differs by less than 3\% from that without sorting. In the NXBG subtraction, we adjusted the pulse height of the calibration sources and instrumental lines (see Koyama et al. 2006a) in the NXBG data to those of the GCDX. The NXBG-subtracted spectrum of the three FI CCDs is given in figure \ref{fig:hardspec}. The back-illuminated (BI) CCD spectrum was constructed  in the same manner.

We can see many line structures in the 6--9~keV band as well as an apparently featureless continuum emission in the 9--11.5~keV band. 
In order to obtain the center energy, flux and width of each line, we simultaneously fit the spectra from the three FI CCDs and the BI CCD spanning the 5.5--11.5~keV band. The model used here is bremsstrahlung absorbed by $N_{\rm Fe}$ for the continuum, adding Gaussians for emission lines. 
In the case of a line blend without sufficient counts to unambiguously resolve all the components, we fixed the energies of 
weaker lines relative to the stronger lines, referring the atomic data from Kaastra and Mewe (1993).
For example, the line energy of Fe\emissiontype{I}~K$\beta$ is fixed to 1.103~$\times E$~(Fe\emissiontype{I}~K$\alpha$), that of Fe\emissiontype{XXV}~K$\beta$ is $E$~(Ni\emissiontype{XXVII}~K$\alpha$)+110~eV, and that of 
Fe\emissiontype{XXV}~K$\gamma$ is $E$~(Fe{XXVI}~Ly$\beta$)+44~eV. 
As for the widths of weak lines, they were either fixed to 30~eV (see next section) or fixed to the likely values referring to the related lines (see table\ref{tabl:lines}). The best-fit model is given in figure \ref{fig:hardspec}, while the best-fit parameters are given in table \ref{tabl:lines}.

 \begin{table*}[!ht]
     \begin{center}
     \caption{The best-fit parameters of the GCDX. }
     \label{tabl:lines}
 \begin{tabular}{lcccc}
       \hline\hline
       \multicolumn{5}{c}{Continuum}\\
       \hline
       $N_{\rm H}$ (cm$^{-2}$)  &&&&6$\times 10^{22}$ (fixed)\\
       $N_{\rm Fe}$ (cm$^{-2}$) &&&&9.7$^{+0.7}_{-0.4}\times 10^{18}$\\
       $kT_{\rm e}$ (keV)&&&&15$^{+2}_{-1}$\\
       \hline
       \multicolumn{5}{c}{Emission lines}\\
       \hline
       Center Energy  &   \multicolumn{2}{c}{Identification}& Width$^b$        & Intensity \\
         (eV)             &         Line    ~      & 
Energy (eV)$^a$        ~ & (eV)  &(photons~s$^{-1}$~cm$^{-2}$)\\
       \hline
       6409$\pm$1  &  Fe\emissiontype{I}$^\dag$~K$\alpha$&                 & 33$^{+2}_{-4}$ & 4.32$^{+0.05}_{-0.08}\times 10^{-4}$\\
       6680$\pm$1  &  Fe\emissiontype{XXV}~K$\alpha$     & 6637--6701 & 39$\pm$2 & 5.10$^{+0.08}_{-0.06}\times 10^{-4}$\\
       6969$^{+6}_{-3}$  &  Fe\emissiontype{XXVI}~Ly$\alpha$   & 6966          & 15$^{+8}_{-15}$  & 1.66$^{+0.09}_{-0.11}\times 10^{-4}$\\
       7069$^*$              &  Fe\emissiontype{I}$^\dag$~K$\beta$ &                 & 38$^\S$            & 6.91$^{+1.12}_{-0.96}\times 10^{-5}$\\
       7490$^{+12}_{-14}$    &  Ni\emissiontype{I}$^\dag$~K$\alpha$&                 & 0 ($<$28)          & 3.05$^{+0.73}_{-0.57}\times 10^{-5}$\\
       7781$^{+24}_{-31}$      &  Ni\emissiontype{XXVII}~K$\alpha$   & 7735--7805     & 39$^|$             & 3.97$^{+1.06}_{-0.65}\times 10^{-5}$\\
       7891$^\#$               &  Fe\emissiontype{XXV}~K$\beta$      & 7881            & 30 (fixed)         & 4.69$^{+0.81}_{-0.61}\times 10^{-5}$\\
       8220$^{+31}_{-22}$      &  Fe\emissiontype{XXVI}~Ly$\beta$     & 8251            & 30 (fixed)         & 2.29$^{+1.35}_{-1.31}\times 10^{-5}$\\
       8264$^{**}$             &  Fe\emissiontype{XXV}~K$\gamma$     & 8295            & 30 (fixed)         & 3.08$^{+1.32}_{-1.34}\times 10^{-5}$\\
       8681$^{+33}_{-32}$      &  Fe\emissiontype{XXVI}~Ly$\gamma$   & 8700            & 0 ($<$91)          & 1.77$^{+0.62}_{-0.56}\times 10^{-5}$\\
       \hline
       Calibration source line&&&&\\
       5896$^{+1}_{-1}$   &  Mn\emissiontype{I}~K$\alpha$      &  5895          & 31$^{+1}_{-1}$ &\\
       6489$^{+1}_{-2}$   &  Mn\emissiontype{I}~K$\beta$       &  6490          & 38$^{+1}_{-3}$ &\\
       \hline
     \end{tabular}
\\
   The errors are at 90\% confidence level.\\
$^a$ Chandra ATOMDB1.3
(http://cxc.harvard.edu/atomdb/WebGUIDE/index.html) \\
and Wargelin et~al. (2005).\\
   $^b$ one standard deviation (1 $\sigma$). \\
   $^*$ fixed to 1.103~$\times E$ (Fe\emissiontype{XXV}~K$\alpha$).\\
   $^\dag$ neutral or low ionization state.\\
   $^\S$ fixed to 1.103~$\times~\sigma$ (Fe\emissiontype{XXV}~K$\alpha$).\\
   $^|$ fixed to $\sigma$ (Fe\emissiontype{XXV}~K$\alpha$).\\
   $^\#$ fixed to $110~+ E$ (Ni\emissiontype{XXVII}~K$\alpha$).\\
   $^{**}$ fixed to 44$~+E$ (Fe\emissiontype{XXVI}~Ly$\beta$).\\
 \end{center}
 \end{table*} 

\subsection{The Reliability of the Center Energy and Width of the Emission Lines}

Since the Fe\emissiontype{XXVI}~Ly$\alpha$ line has a relatively simple structure, its centroid can be accurately predicted. 
In the case of our measurement, the line flux is strong enough to determine the centroid energy with high accuracy. 
The theoretically predicted and observed centroids are  6966eV and  6969$^{+6}_{-3}$ eV
(errors in this paper are at 90\% confidence level unless otherwise mentioned). Also the calibration energy of the characteristic K$\alpha$ and K$\beta$ lines from neutral Mn (Mn\emissiontype{I}) agree 
with the theoretical energy within 1 eV. These agreements demonstrate that our procedure of the CTI correction and fine gain-tuning worked well. 
Since the error on the center energy of Fe\emissiontype{XXVI}~Ly$\alpha$ is  $^{+6}_{-3}$ eV, we regard the over-all systematic error of our gain determination in the iron energy band to be within $^{+3}_{-6}$ eV.

The observed line widths (after the de-convolution of the response function) 
of Mn\emissiontype{I}~K$\alpha$ and Fe\emissiontype{I}~K$\alpha$ are 31$\pm$1~eV and  33$^{+2}_{-4}$~eV (1 $\sigma$), respectively. 
The energy resolution has been gradually decreasing since the launch in July 2005, and the line broadening is consistent with the long-term CCD degradation for the calibration line.
The time dependent energy resolution is currently not well-enough understood to implement here, but it is important to note that
the observations reported here were made early in the mission.
The apparent line broadening (here, the systematic broadening) of $\sim$30eV at $\sim$6--7 keV is mainly due to the long-term degradation of the CCD 
energy resolution, since, as described below, there are internal consistencies that argue against other systematic causes.  
(1) The calibration line ($^{55}$Fe) is irradiating a small spot of the CCD, therefore any line broadening due to  non-uniformity of the gain and non-linearity of the CTI (if any) in the CCD area can be ignored for the calibration line.
(2) Because the 33 eV line broadening of the 6.4 keV line in the GCDX spectrum agrees with the broadening 
of the calibration line (31 eV at 5.9 keV), it follows that the source of the 
broadening is the same for both the calibration and the GCDX line feature.
(3) The observed line broadening for the 6.9 keV line is less than 
the calibration line (see discussion below) 
, hence  no line broadening is found in the 6.9 keV line,
(4) We divided the observations into two terms (Seq No. 100027010+100027020, and 100037010+100037040, see table 1) separated by about one 
week and examined the center energies.
The observed difference of the center energy of Fe\emissiontype{XXV}~K$\alpha$ is only 1~eV, well within the statistical error of 1~eV. Therefore the gain is almost constant in the observational period, hence does not cause  any line broadening in the integrated spectrum.

Apart from the line broadening caused by the long-term degradation of the CCDs, 
we can safely conclude that the response function is well-calibrated for the present observation, especially for the line center energy. 

\section{Discussion} 

\subsection{Over Abundance of Fe in the Interstellar Gas}

The continuum shape of the GCDX spectrum is well reproduced with the addition of 
an Fe K-edge at 7.1~keV. The best-fit column density, $N_{\rm Fe}$, is determined to be 
9.7$^{+0.7}_{-0.4}\times 10^{18}$~cm$^{-2}$. This corresponds to an iron abundance 
of 3.5 times solar assuming a typical hydrogen column density ($N_{\rm H}$) toward the GC of
6$\times 10^{22}$~cm$^{-2}$ \citep{Sa02} and the solar abundances of  Anders and  Grevesse (1993).
We expect that a large fraction of this absorption takes place in the dense gas clouds 
prevailing near the GC region. 

\subsection{K$\alpha$ Line from Fe\emissiontype{I}}

The energy of Fe\emissiontype{I}~K$\alpha$ is 6400~eV (Kaastra and Mewe 1993), which is lower than that found in the 
GCDX of 6409$\pm$1~eV by 9~eV.  However, because the energy difference of 9 eV is larger than the systematic error 
of $^{+3}_{-6}$ eV, we suspect that this difference is a result of blending of Fe\emissiontype{I}~K$\alpha$ 
with line emission from low charge states of iron. The line width of 33$^{+2}_{-4}$~eV is comparable to, but slightly larger than,  the systematic broadening of $\sim$30~eV, which may also be the result of
the line blending.\\

\subsection {K$\alpha$ Line from Fe\emissiontype{XXVI}}

The line width of Fe\emissiontype{XXVI}~K$\alpha$ is determined to be $\le$23~eV, which is smaller than 
the systematic broadening of$\sim$30 eV . This unnaturally narrow line could be due to the coupling with the nearby line Fe\emissiontype{I}~K$\beta$. 
We therefore examine the coupling effect by fixing to the theoretically 
predicted flux of 0.125$\times$  K$\alpha$  (Kaastra and Mewe 1993). The best-fit  result of 25$^{+5}_{-8}$ eV is consistent with the  systematic broadening.  
Thus we can safely conclude that the line width broadening of Fe\emissiontype{XXVI}~K$\alpha$ is instrumental.

\subsection{The Center Energy and Width of Fe\emissiontype{XXV}~K$\alpha$}

The Fe\emissiontype{XXV}~K$\alpha$ line is a blend of the resonance, inter-combination, and forbidden lines from Fe\emissiontype{XXV} (see table \ref{tbl:tripletline}), and depending on the conditions of the source plasma, may also contain satellite lines produced by dielectronic recombination and innershell excitation \citep{Be92,Be93}. The fact that the observed line centroid and  width of Fe\emissiontype{XXV}~K$\alpha$ depends on the flux ratio of these lines, which, in turn, are a function of the  plasma conditions, make the K$\alpha$ line feature an excellent diagnostic. As discussed above, a centroid shifted towards lower energy relative to the CIE case is an indication of a recombining  plasma. For the GCDX, the observed center energy of Fe\emissiontype{XXV}~K$\alpha$ is 6680$\pm$1~eV. 
We compare our result to the laboratory values produced by the CX recombination, i.e., a recombination dominated plasma, and also to calculations using APEC and MEKAL of the line emission produced by a CIE plasma. The laboratory measured line centroid of Fe\emissiontype{XXV}~K$\alpha$ produced by the CX recombination alone, is $6666\pm 5$ eV \citep{Wa05}, significantly lower than the value measure in the GCDX. The APEC and MEKAL predicted centroid for the CIE case are 6685~eV and 6680~eV, respectively, in agreement with the GCDX value of 6680~eV. 

We also note that the line broadening of 39$\pm$2~eV for Fe\emissiontype{XXV}~K$\alpha$  is larger than the systematic broadening of $\sim$30 eV (see section 3.2). This must be due to the fact that several lines contribute to this line feature. 

   \begin{table}[!ht]
     \begin{center}
     \caption{Table of He-like Iron triplets}
    \label{tbl:tripletline}
    \begin{tabular}{lll}
         \hline\hline
	 Name & Transition & Energy (eV)$^a$ \\
	 \hline
	 Resonance        &$1s^2$ $^1S_0$--$1s2p$ $^1P_1$&6701\\
	 Intercombination &$1s^2$ $^1S_0$--$1s2p$ $^3P_1$&6682\\
	 Intercombination &$1s^2$ $^1S_0$--$1s2p$ $^3P_2$&6668\\
	 Forbidden        &$1s^2$ $^1S_0$--$1s2s$ $^3S_1$&6637\\
         \hline \\
       \end{tabular}\\
$^a$ from Wargelin et~al. (2005).\\
     \end{center}
   \end{table}

\subsection{Constraints on the Cosmic Ray Iron Velocity in Charge Exchange Scenario} 
The relatively high resolution of the XIS also affords us the ability to extract crucial information from the line feature of Fe\emissiontype{XXVI}~Ly$\alpha$ provides further evidence 
against the CX process for X-ray production.  As discussed in section 4.3,  the intrinsic line broadening  of Ly$\alpha$ is nearly zero, and it follows that if CX is the source of the X-ray emission, then recombination would occur  when the bare iron  velocity is nearly zero, i.e., the collision energy between the iron ion and neutral hydrogen is small. In such a low-collision energy case, electrons are non-statistically captured into levels with large principal quantum numbers ($n\geq10$) \citep{Wa05,Pe01,Ot06}.  We, however, see no enhancement at the Fe\emissiontype{XXVI}~Rydberg series limit at 9.2~keV. Our measurements gives an upper limit of $9\times 10^{-6}$~photons~s$^{-1}$~cm$^{-2}$ for the high-$n$ Rydberg series line emission (we assumed a Gaussian line with width of 30~eV), only about 6\% of the intensity of Ly$\alpha$. This upper limit corresponds to a lower limit of collision energy of 100~eV amu$^{-1}$ (\cite{Pe01};  \cite{Wa05}) between bare Fe nuclei and neutral H. For that collision energy, the velocity of bare iron nuclei must be larger than $\sim$150~km~s$^{-1}$. Therefore the CX process, if present in the GCDX, gives a lower limit on the velocity of $\sim$150~km~s$^{-1}$.
\citet{Ta02} suggested that the CX process may have the maximum cross section at a higher iron velocity of $\sim 5000$km s$^{-1}$, hence the subsequent decay lines have widths of about 100~eV. 
No such large broadening is observed in Fe\emissiontype{XXVI}~Ly$\alpha$.

Together with the discussion of section 4.4, we therefore propose that the 6.7~keV and 6.9~keV lines in the GCDX are likely due to a CIE plasma and would not arise from a CX process.

\subsection{The Flux Ratio of K-shell Transition Lines from Fe\emissiontype{XXV, XXVI} and Ni\emissiontype{XXVII}}

The K-shell lines of Fe\emissiontype{XXV}~K$\alpha$ and 
Fe\emissiontype{XXVI}~Ly$\alpha$ were discovered with the data from ASCA~\citep{Ko96}, Chandra~\citep{Mu04}  and XMM-Newton~\citep{Pr03}. 
With Suzaku, we have, for the first time, detected Ni\emissiontype{XXVII}~K$\alpha$, and Fe\emissiontype{XXV}~K$\beta$ and Fe\emissiontype{XXVI}~Ly$\beta$. 
We have also tentatively identified  Fe\emissiontype{XXV}~K$\gamma$  and  Fe\emissiontype{XXVI}~Ly$\gamma$ 
(see table 2).   
As discussed in sections 4.4 and 4.5, the emission lines from these highly ionized atoms are likely due to collisional excitation  in a thin thermal plasma. 
The best-fit flux ratio of 
Fe\emissiontype{XXVI}-Ly$\alpha$/Fe\emissiontype{XXV}-K$\alpha$ is 0.33$\pm$0.02, which provides an ionization temperature of  6.4$\pm$0.2~keV. The best-fit flux ratio of Fe\emissiontype{XXV}-K$\beta$/Fe\emissiontype{XXV}-K$\alpha$ is 0.092$^{+0.016}_{-0.012}$, which gives the electron temperature of 6.6$^{+4.2}_{-1.7}$~keV. 
Although the error is still large, this electron temperature is similar to the ionization temperature. 
The line flux ratio of 
Ni\emissiontype{XXVII}-K$\alpha$/Fe\emissiontype{XXV}-K$\alpha$ is 0.078. With the further assumption that the relative abundances of iron and nickel are proportional to the solar value, this ratio is converted  to the plasma temperature of $\sim$5.4~keV. We therefore conclude that the GCDX plasma has a temperature of $\sim$5--7 keV in collisional ionization equilibrium.
The observed flux ratios of 
Fe\emissiontype{XXVI}-Ly$\beta$/Fe\emissiontype{XXVI}-Ly$\alpha$,
 Fe\emissiontype{XXV}-K$\gamma$/Fe\emissiontype{XXV}-K$\alpha$ fr
 and Fe\emissiontype{XXVI}-Ly$\gamma$/Fe\emissiontype{XXVI}-Ly$\alpha$ from are consistent with this plasma condition. 

\subsection{Temperature and Flux Variations of the GCDX Plasma}

\begin{figure}[!ht]
   \begin{center}
     \FigureFile(70mm,50mm){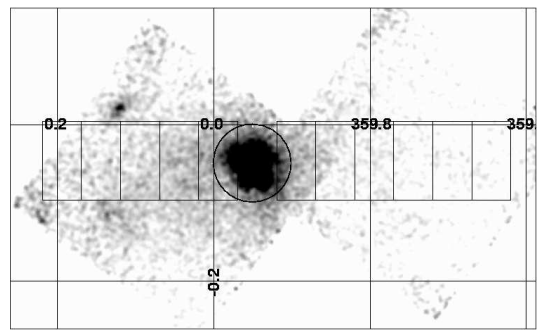} 
     \caption{Image of the 2 fields of view (FOV) in the 6.62--6.74~keV band, overlaid with 
the rectangular regions from which the spectra were extracted.}
     \label{fig:boxreg}
   \end{center}
 \end{figure}

\begin{figure}[!ht]
   \begin{center}
    \FigureFile(70mm,50mm){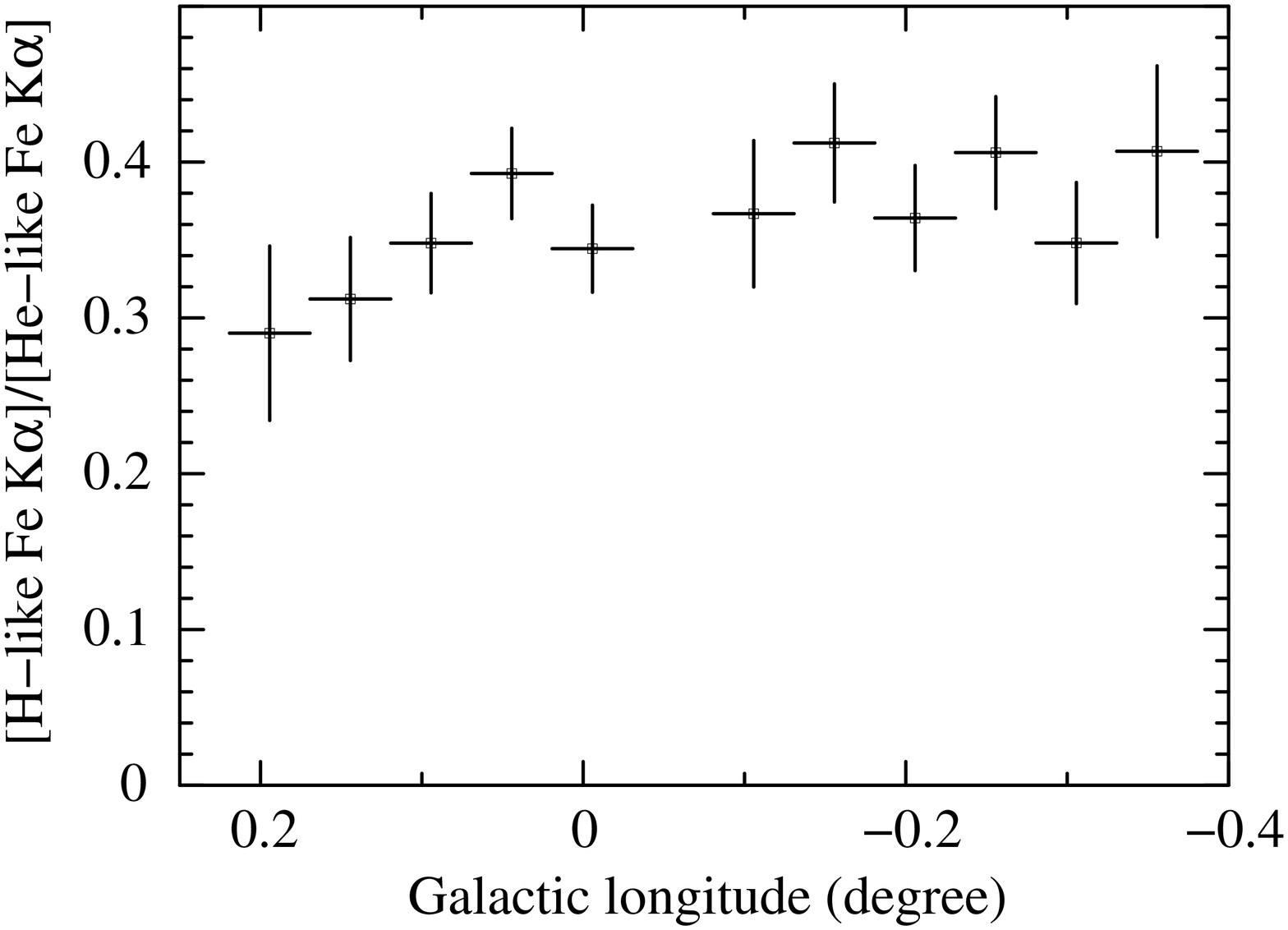} 
     \caption{The line flux ratios of Fe\emissiontype{XXVI}-Ly$\alpha$/Fe\emissiontype{XXV}-K$\alpha$  of measured in the series of rectangular regions
along the Galactic plane  given in figure 4 (the constant  $b=-\timeform{0D.05}$ line).}
     \label{fig:ratio}
   \end{center}
 \end{figure}

\begin{figure}[!ht]
   \begin{center}
     \FigureFile(70mm,50mm){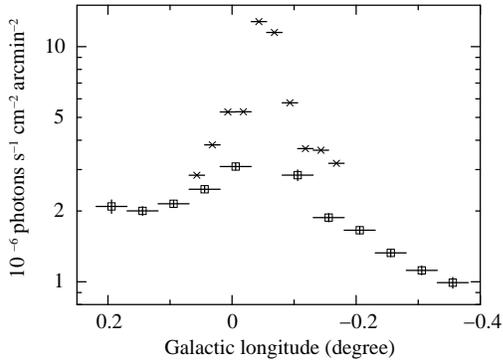} 
 \caption{The line fluxes of Fe{XXV} K$\alpha$ are given by the squares, 
while the integrated point source fluxes in the 4.7--8~keV band are plotted by the crosses. 
The horizontal axis is the same as figure \ref{fig:ratio}, but the vertical axis is a logarithmic scale.}
     \label{fig:ps_diff}
   \end{center}
 \end{figure}

Although we show that the GCDX  is attributable to a high temperature plasma of  $\sim$ 5--7 keV in collisional ionization equilibrium, whether the origin is diffuse or the integrated  emission of a large number of unresolved point sources, is not yet clear.  We therefore investigated the spatial variation of the GCDX in temperature and flux. We made a series of X-ray spectra in $3\times6$ arc-min rectangles along the Galactic plane of $b=-\timeform{0D.05}$ (figure \ref{fig:boxreg}), then extracted  the fluxes of the 6.7 and 6.9~keV lines, applying the same model as is given in table \ref{tabl:lines}. The plots of the flux ratio between the 6.7 and 6.9~keV lines, and the fluxes of the 6.7 keV line are given in figure \ref{fig:ratio} and figure \ref{fig:ps_diff}, respectively, in which the flux and flux ratio of the bright SNR Sgr A East are excluded. We see that the 6.7 and 6.9~keV line ratio is constant at $\sim0.38$ in the negative longitude region from $l= -\timeform{0D.1}$ to $l=-\timeform{0D.4}$, but decreases to $\sim0.30$ at the positive longitude near at $l = \timeform{0D.25}$. The ratios of 0.38 and 0.30 
suggest ionization temperatures of 6.8~keV and 6.2~keV, respectively. The smooth and monotonic variation of the ionization temperature along the 
galactic longitude argues that the GCDX is not due to the superposition of the 
outputs of a large number of unresolved point sources.

We simulated the point-source flux in the 4.7--8.0~keV band observed with XIS, putting the positions and fluxes (2.0--8.0~keV band) of the  Chandra cataloged sources \citep{Mu03} in the same rectangles of figure 4.  
To preserve completeness limit, we selected the sources with flux larger 
than 5$\times 10^{-7}$ photons~cm$^{-2}$~s$^{-1}$ in the 2.0--8.0~keV band.  After convolving the response of the XRT + XIS, we extracted the integrated point-source fluxes from each rectangle.  
The results are plotted in the same figure of the 6.7 keV line (figure \ref{fig:ps_diff}). 
The integrated point-source flux is symmetrically distributed between positive and negative galactic longitude with a sharp peak at Sgr A$^*$  ($l = -\timeform{0D.06}$), which is significantly different from the 6.7 keV line flux distribution. 
The 6.7 keV line flux  shows an asymmetric distribution (larger flux at the positive galactic longitude than the negative side) and a gradual increase towards Sgr A$^*$.
These facts suggest that the major fraction of the 6.7 keV line flux is not due to the integrated emission of unresolved point sources, but is truly diffuse.  

\subsection{Hard X-ray Tail}

\begin{figure}[!ht]
   \begin{center}
  \FigureFile(70mm,50mm){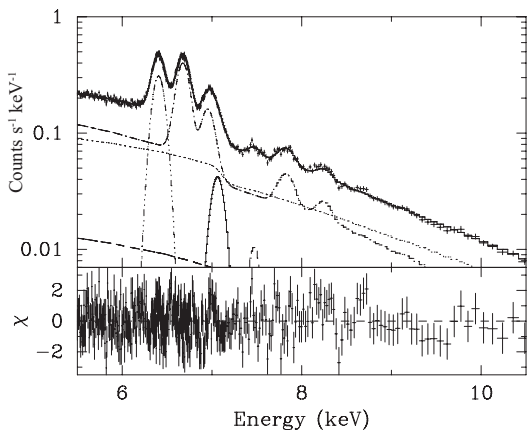} 
     \caption{Same as figure \ref{fig:hardspec}, but the model is a CIE plus  a power-law with 3 Gaussian lines and an iron absorption edge.}
     \label{fig:apecfit}
   \end{center}
 \end{figure}

From the line diagnostics, the GCDX is most likely produced by a thin hot plasma with a temperature of $\sim$5--7~keV in 
CIE, while the bremsstrahlung temperature determined with the continuum shape in the 5.5--11.5~keV band is $\sim$15~keV(see table 2), significantly higher than $\sim$5--7~keV. This suggests that the
continuum flux may contain an additional hard component.
We therefore fit the spectrum with a model of a CIE plasma and  power-law component with 3 Gaussians for the K-shell emissions from Fe\emissiontype{I} and Ni\emissiontype{I}. In this model, the K-edge absorption with $N_{\rm Fe}$ is also implemented. Although its contribution is minor, we added the model spectrum of the cosmic X-ray background (CXB).
As discussed in section 3.2, there are systematic errors in the line width ($\sim$30 eV) and center energy 
($^{+3}_{-6}$~eV).  
We therefore introduced an artificial random motion and small red-shift to cancel these systematic errors.  
The fit is reasonably good and the best-fit results and parameters are given in figure \ref{fig:apecfit} 
and table \ref{tabl:vapec}.
We note that the forbidden line of Ni\emissiontype{XVII} and some of the satellite lines are 
missing in the APEC model, hence the Ni abundance may be overestimated. 

\begin{table*}
  \begin{center}
    \caption{Best-fit parameter with the VAPEC model}
     \label{tabl:vapec}
\begin{tabular}{lcccc}
      \\
      \hline\hline
      Parameter & Unit   &&             &  Value\\
      \hline
      $N_{\rm H}$       & cm$^{-2}$ &&& 6$\times 10^{22}$ (fixed)\\
      $N_{\rm Fe}$      & cm$^{-2}$ &&& 8.5$^{+0.8}_{-0.5}\times 10^{18}$ \\
      $kT_{\rm e}$    & keV            &&     & 6.5$^{+0.1}_{-0.1}$\\
      $Z_{\rm Fe}$ & solar             &&  &1.2$^{+0.5}_{-0.3}$ \\
      $Z_{\rm Ni}$ & solar             &&  &2.1$^{+0.7}_{-0.5}$ \\
      Velocity     & km~s$^{-1}$ &&&1192\\
      Redshift    &                    && &6.0$\times 10^{-4}$\\
      \hline
\multicolumn{5}{c}{Hard Tail}\\
\hline
       $\Gamma$ & & & & 1.4$^{+0.5}_{-0.7}$ \\
       Flux at 8 ~keV & photons~s$^{-1}$~cm$^{-2}~$keV$^{-1}$ & &
&4.5$^{+1.9}_{-2.1}\times 10^{-4}$ \\
      \hline
\multicolumn{5}{c}{Emission lines}\\
       \hline
       Center Energy  &   \multicolumn{2}{c}{Identification}& Width
   & Intensity \\
         (eV)             &         Line    ~        &  Energy (eV)
    ~ & (eV)  &(photons~s$^{-1}$~cm$^{-2}$)\\
       \hline
       6409$^{+1}_{-1}$  &  Fe\emissiontype{I}$^\dag$~K$\alpha$&
               & 36$^{+1}_{-3}$ & 4.41$^{+0.05}_{-0.08}\times 10^{-4}$\\
       7069.1$^*$  &  Fe\emissiontype{I}$^\dag$~K$\beta$ &
   &39$^*$&7.8$^{+0.4}_{-0.6}\times 10^{-5}$\\
       7492$^{+14}_{-18}$&  Ni\emissiontype{I}$^\dag$~K$\alpha$&   &
0($<$30) &2.3$^{+0.7}_{-0.4}\times 10^{-5}$\\
      \hline
    \end{tabular}
\\
   $^\dag$ or low ionization state\\
   $^*$ fixed to 1.103~$\times E$ (Fe\emissiontype{XXV}~K$\alpha$) .\\

  \end{center}
  \vspace{-0.5cm}
 \end{table*}

One notable finding is the presence of power-law hard tail. The power-law index $\Gamma$ is sensitive to the slope of the high energy continuum.
Because the effective area of the XRT-XIS combination drops rapidly at high energy,
our determination of the power-law index $\Gamma$ has a relatively large statistical 
error  of  $\Gamma\sim$1.4$^{+0.5}_{-0.7}$.
The low numbers of counts make it challenging to determine the error contribution from systematic effects. To provide a more reliable estimate of the systematic error, we checked it using the 
spectrum of the Crab nebula. We examined the variation of the best-fit $\Gamma$ by fitting the Crab 
spectrum in the 
4--10, 5--10, 6--10 and 7--10~keV bands, and found no systematic variations with the 
increasing energy band. The mean variations of $\Gamma$ are within 5\% for all the XISs, which is far smaller than the 
statistical error of $\sim$40\%.

Since the flux level of the NXBG becomes comparable to the hard tail flux at 8--10 keV, it may be argued that the error due to the NXBG-subtraction should not be ignored. 
The possible flux error of the NXBG is estimated to be $\sim\pm$3\% (Koyama et al. 2006a). 
Including  this error in the spectrum, the best-fit $\Gamma$ are 1.4 (adding 3\% of the NXBG) and 1.2 (3\% reduction of the NXBG). 
Further systematic analysis of the NXBG found a hint that the flux  of the 
night Earth data is smaller than that from a blank sky.  The details of the flux difference is still in 
study, but is at a level of $\leq$ 10\% at the 8--11 keV band. If we use a NXBG of 1.1$\times$ of the night Earth,
then the  best-fit $\Gamma$ becomes 1.5.
Thus any possible systematic error of NXBG does not change the conclusion that a flat hard tail is 
present in the GCDX. 
In addition, the flux of the CXB is more than  one order of magnitude smaller than the hard tail in the hard X-ray band,
so that the fluctuation and/or error of the CXB can be ignored.  

A possible contributor of the power-law hard tail is non-thermal filament structures found  
with Chandra  (\cite{Mo02}; \cite{Se02}; \cite{Pa04}). 
In fact, the power-law flux in the positive galactic longitude (GC\_SR1) is significantly larger than that of the negative galactic longitude (GC\_SR2), which is a trend similar to the population  
of the non-thermal filaments (\cite{Mo02};  \cite{Se02}; \cite{Pa04}).
The  origin of the non-thermal emission is a key for understanding the presence of high energy particles, 
which are suggested by the detection of very high energy Gamma-rays (\cite{Ah06}; \cite{Ts04}; \cite{Ko04}). 
Further analysis of the XIS data in combination with those of the Hard X-ray Detector(HXD) (\cite{Ta06}) 
could yield an important contribution to the study of this hard component.

\section{Summary}

\begin{enumerate}

\item Since the difference between the theoretical (6966~eV) and observed (6969$^{+6}_{-3}$~eV)
values of Fe\emissiontype{XXVI}~Ly$\alpha$ is only 3~eV, we can assume that the detector gain and uniformity  are well-calibrated to study features at this level of accuracy.

\item The abundance of  Fe in the interstellar gas near the GC appears to be 3.5 times solar.

\item The 6.4~keV (K$\alpha$ of neutral iron) line from the GC region could be contaminated by
those of low ionization states, but still characteristic of gas at temperature much lower than those in which the helium-like lines arise. 

\item The observed line centroid energy of Fe\emissiontype{XXV}~K$\alpha$ is 6680$\pm$1~eV, 
which is consistent with that of the collisional ionization plasma.  

\item The GCDX spectrum is consistent with that of a CIE plasma with mean temperature 5--7~keV.

\item The plasma temperature of GXDX is constant at 6.8 keV in the galactic 
longitude region from $l=-\timeform{0D.1}$ to $l =-\timeform{0D.4}$, while exhibiting  a significant decrease to 6.2~keV near $l= \timeform{+0D.25} $.

\item The flux distribution  of the 6.7 keV line along the Galactic  plane 
is different from that of the integrated point sources.

\item We found a clear hard tail in the GCDX.

\end{enumerate}

\bigskip
The authors express  sincere thanks  to all the Suzaku team members, especially A. Bamba, H. Uchiyama, H. Yamaguchi, and H. Mori for their comments
and supports.
Y.H. and H.N. are supported by JSPS Research Fellowship for Young Scientists.
This work is supported by the Grants-in-Aid for the 21st Century COE "Center for Diversity and Universality in Physics" and also by Grants-in-Aid (No 18204015) of 
the Ministry of Education, Culture, Sports, Science
and Technology (MEXT).
Work by UC LLNL was performed under the auspices of the U.S. Department of Energy by University of California, Lawrence Livermore National Laboratory under Contract W-7405-Eng-48.

\appendix
\section*{The Internal Calibration of CTI and Gain Tuning}
\begin{figure}[!ht]
   \begin{center}
     \FigureFile(70mm,50mm){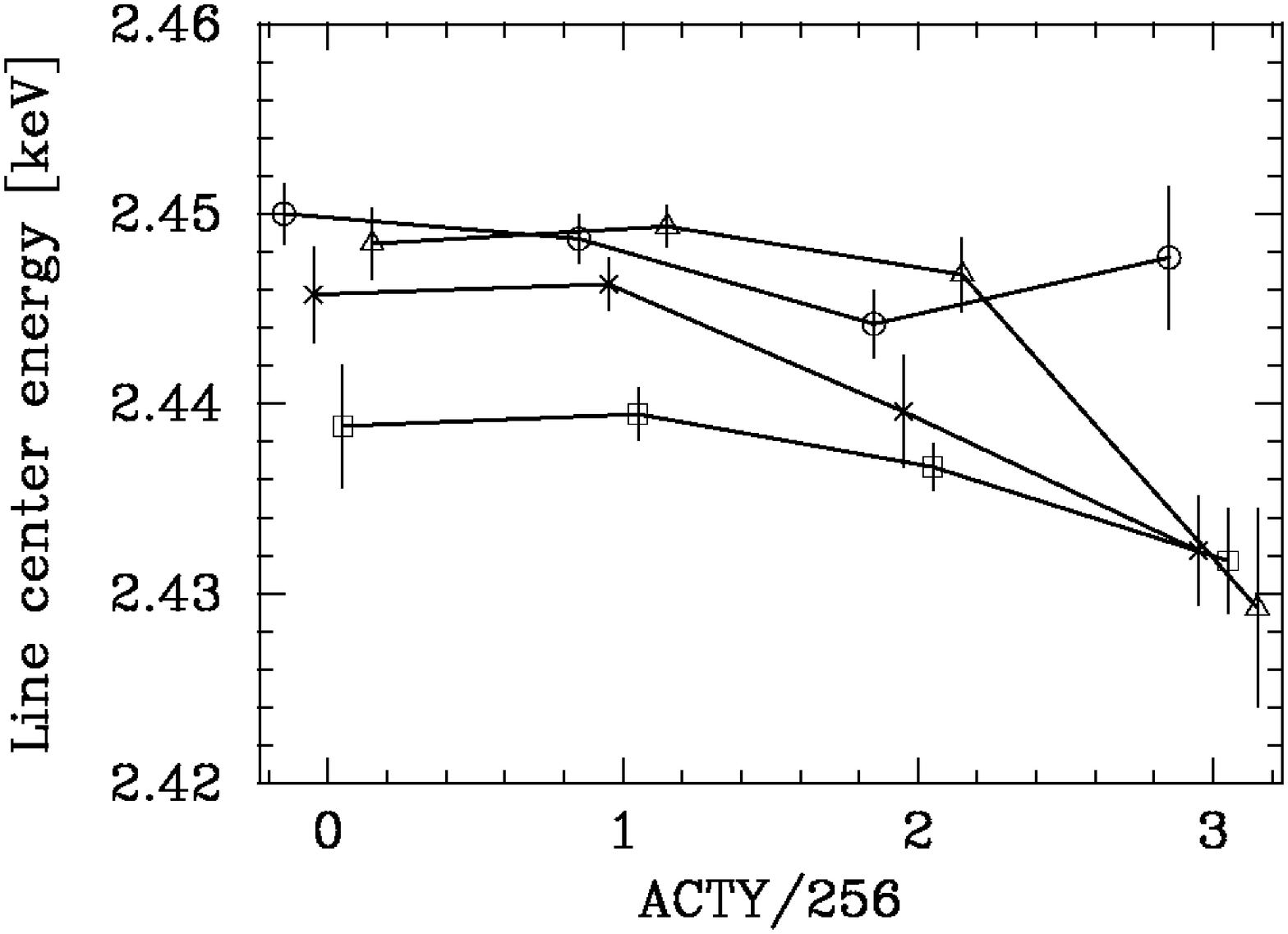} 
    \caption{The center energy of S\emissiontype{XV}~K$\alpha$. In  the 4 regions of IA divided along the ACTY axis for each XIS. Circles, crosses, squares, and triangles represent line center energy obtained with XIS0, XIS1, XIS2, and XIS3 respectively.}
     \label{fig:4xis-cti0Sheka}
   \end{center}

 \end{figure}

\begin{figure}[!ht]
   \begin{center}
     \FigureFile(70mm,50mm){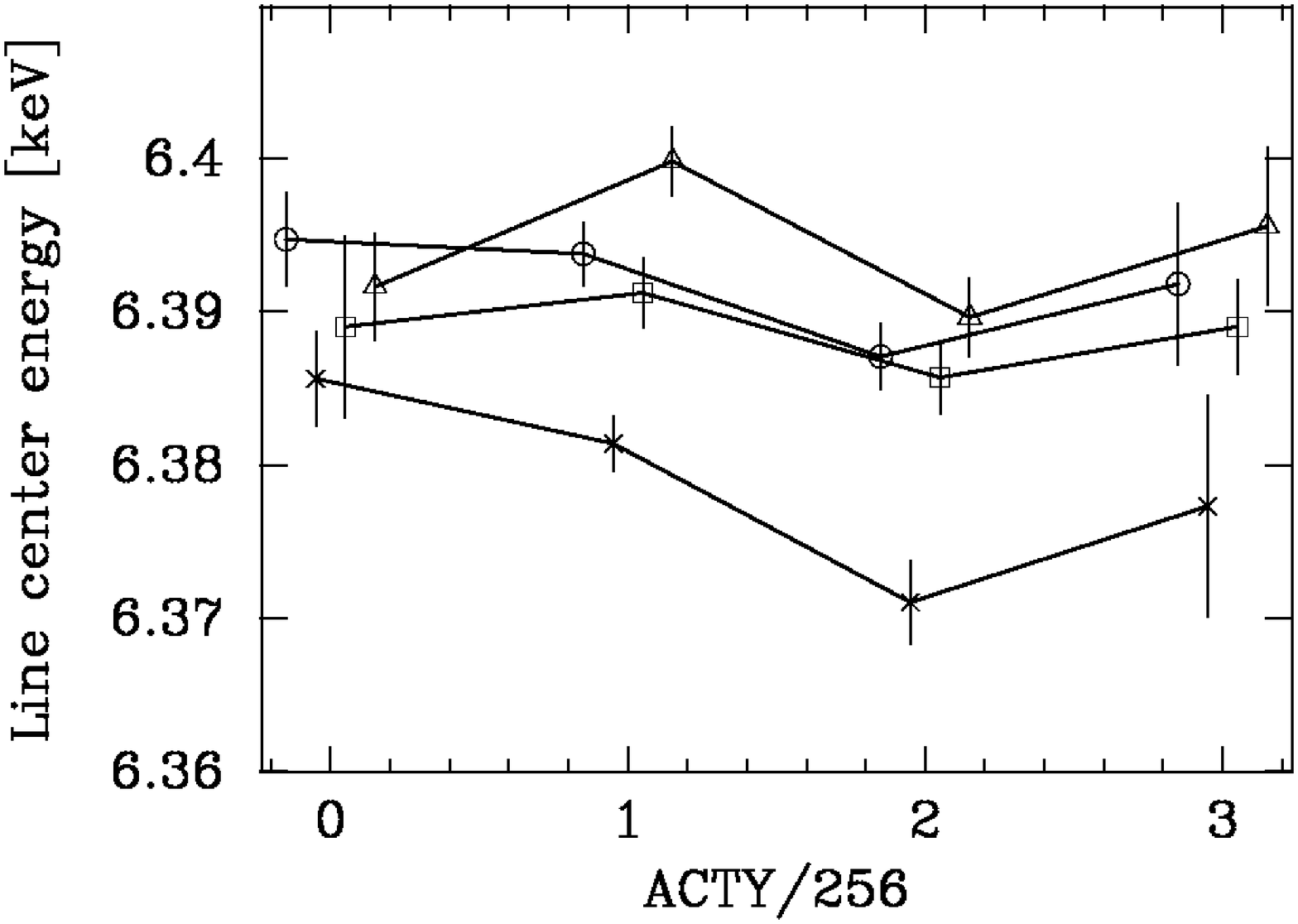} 
     \caption{Same as figure \ref{fig:4xis-cti0Sheka}, but for the Fe\emissiontype{I}~K$\alpha$ line.}
     \label{fig:4xis-cti0Feneuka}
   \end{center}
 \end{figure}

\begin{figure}[!ht]
   \begin{center}
     \FigureFile(70mm,50mm){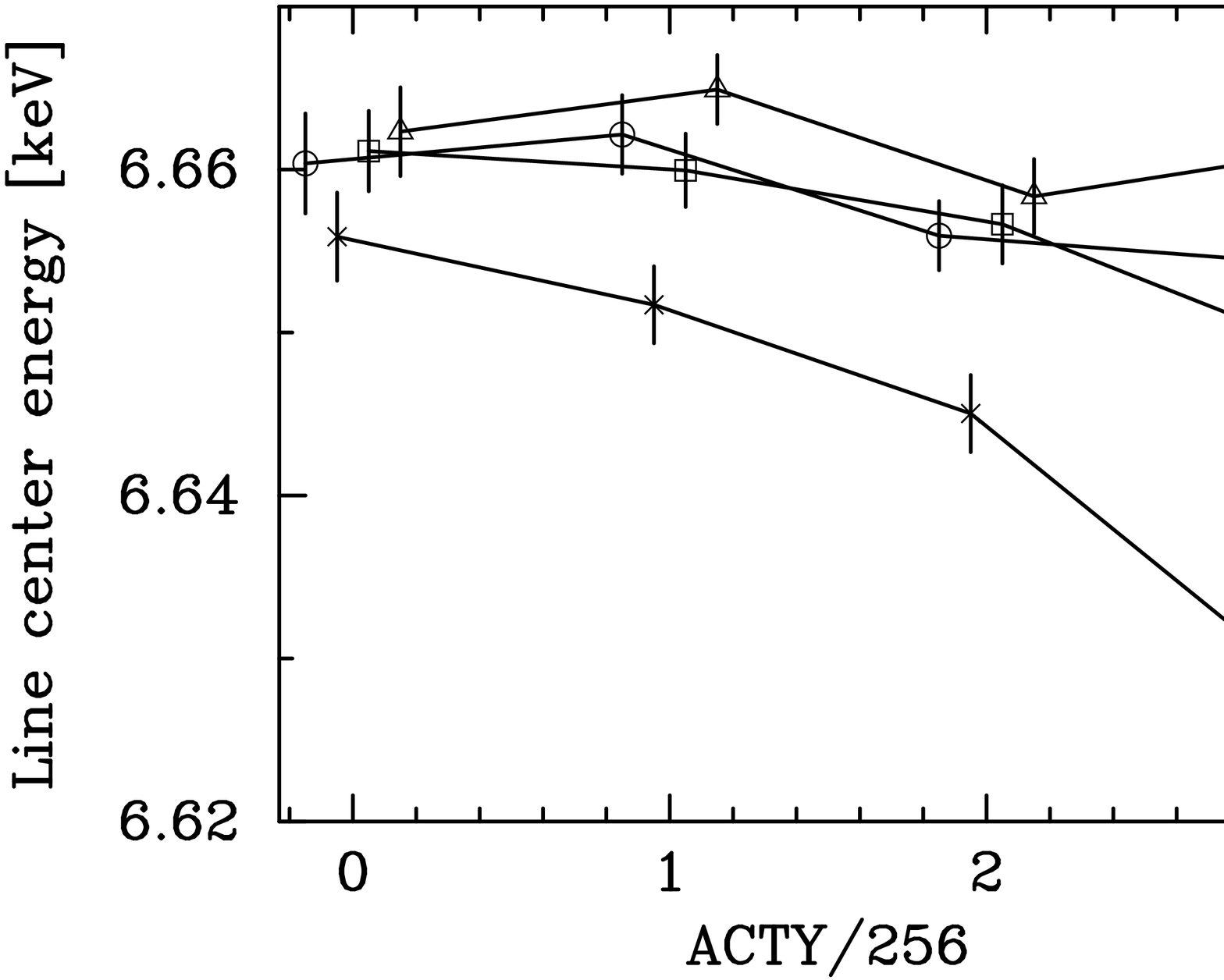} 
     \caption{Same as figure \ref{fig:4xis-cti0Sheka}, but for the of Fe\emissiontype{XXV}~K$\alpha$ line.}
     \label{fig:4xis-cti0Feheka}
   \end{center}
 \end{figure}

Since the GCDX emits strong lines covering over the entire field view of the XIS, we can calibrate the gain and its variation in the full XIS imaging area (IA). For this purpose, we divide each XIS area into 4 sections along the ACTY-axis and designate as No.1 to 4 in the order of the distance from the signal read-out nodes (No.4 corresponds the largest ACTY, while No.1 is the smallest). The definition of ACTY  is given in Koyama et al. (2006a). Figures \ref{fig:4xis-cti0Sheka}, \ref{fig:4xis-cti0Feneuka} and \ref{fig:4xis-cti0Feheka} are the plots of the line center energy of S\emissiontype{XV}~K$\alpha$, Fe\emissiontype{I}~K$\alpha$ and Fe\emissiontype{XXV}~K$\alpha$, respectively.

 
From figures \ref{fig:4xis-cti0Sheka}, \ref{fig:4xis-cti0Feneuka} and \ref{fig:4xis-cti0Feheka}, we see 
a clear trend that  the line center energy is lower at longer distance from the read-out nodes. 
The effect is clearly instrumental (rather than astronomical), since  the trend is observed in all four XIS detectors, with differing read-out orientations on the sky  (Koyama et al. 2006a). 
Therefore, we can safely conclude that the systematic trend that the center energy decreases as 
increasing ACTY is due to the charge transfer inefficiency (CTI). 
Assuming that  CTI is a linear function of the distance from the read-out nodes, we correct the CTI so that the center energies of the three K$\alpha$ lines become constant along the ACTY axis. 
 
\begin{figure}[!ht]
   \begin{center}
     \FigureFile(70mm,50mm){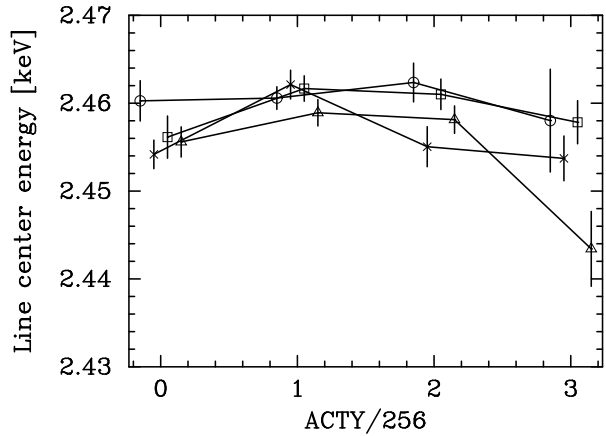} 
     \caption{Same as figure \ref{fig:4xis-cti0Sheka}, but after the CTI correction.}
     \label{fig:4xis-aftctiSheka}
   \end{center}
 \end{figure}
 
\begin{figure}[!ht]
   \begin{center}
     \FigureFile(70mm,50mm){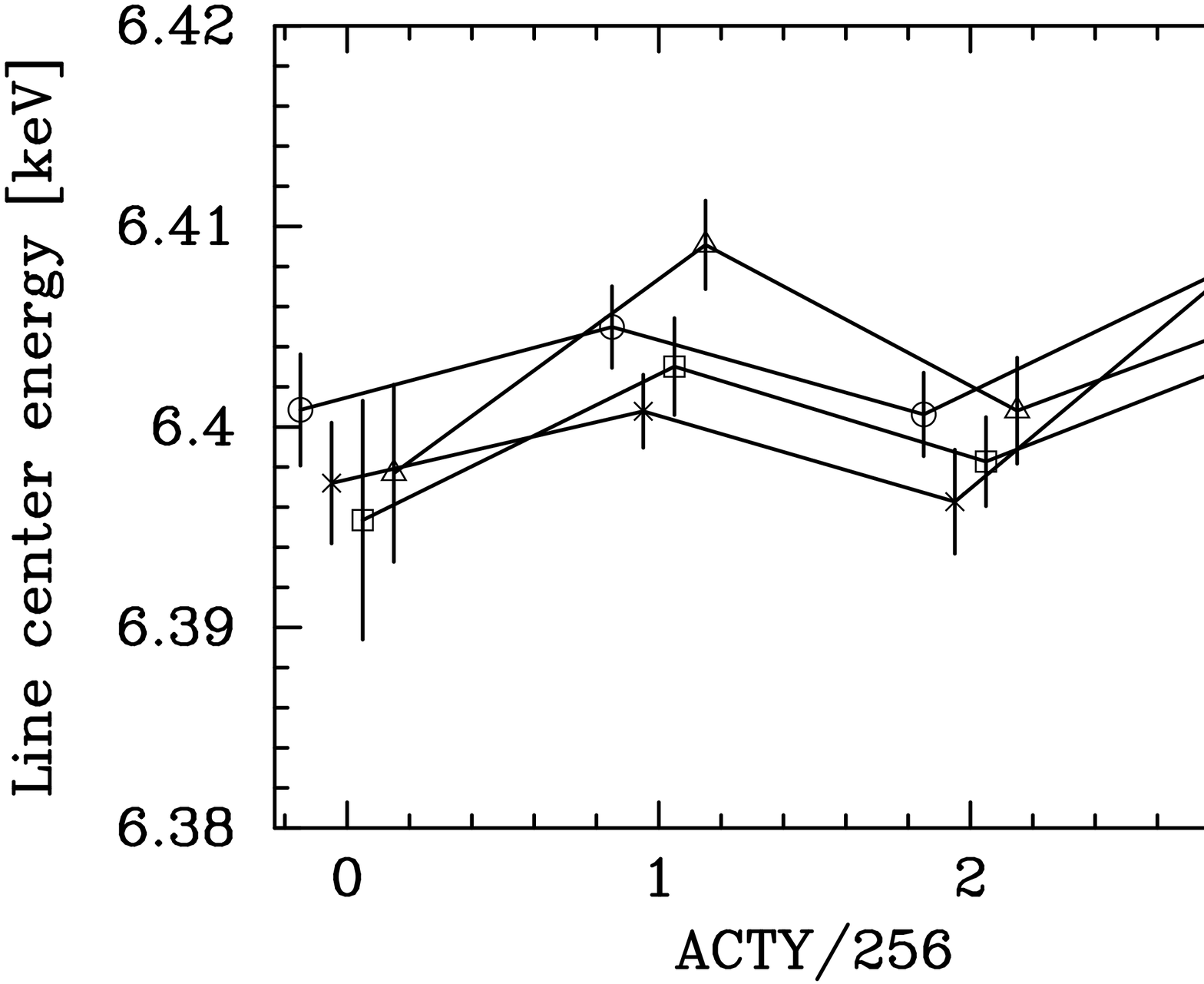} 
     \caption{Same as figure \ref{fig:4xis-aftctiSheka}, but for the  Fe\emissiontype{I}~K$\alpha$ line.}
     \label{fig:4xis-aftctiFeneuka}
   \end{center}
 \end{figure}
 
\begin{figure}[!ht]
   \begin{center}
     \FigureFile(70mm,50mm){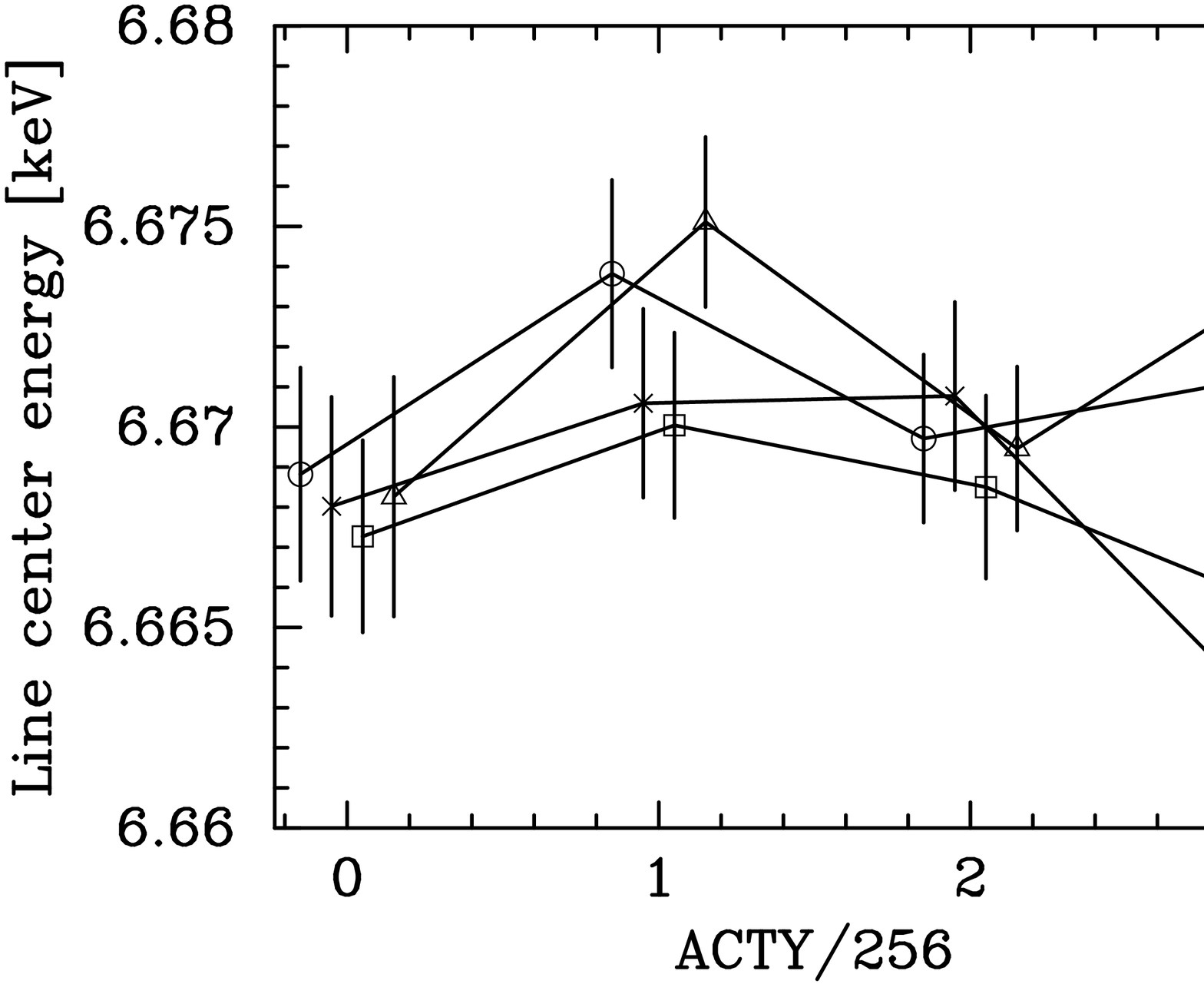} 
     \caption{Same as figure \ref{fig:4xis-aftctiSheka}, but for the  Fe\emissiontype{XXV}~K$\alpha$ line.}
     \label{fig:4xis-aftctiFeheka}
   \end{center}
 \end{figure}

\newpage

The CTI-corrected plots are given in figures \ref{fig:4xis-aftctiSheka}, \ref{fig:4xis-aftctiFeneuka} and \ref{fig:4xis-aftctiFeheka}. The systematic decreases of the center energy along the ACTY-axis disappear, demonstrating
that  the CTI correction is properly made. We should note that the CTI at 6.4 and 6.7 keV is consistent with that at 5.9 keV from  
the calibration sources reported by the XIS team (Koyama et al. 2006a). Note also that we have corrected the CTI for each XIS 
independently, but the CTI is assumed to be identical in the  4-segments. Also absolute energy gain tuning is 
not yet made at this stage. 

\begin{figure}[!ht]
   \begin{center}
     \FigureFile(70mm,50mm){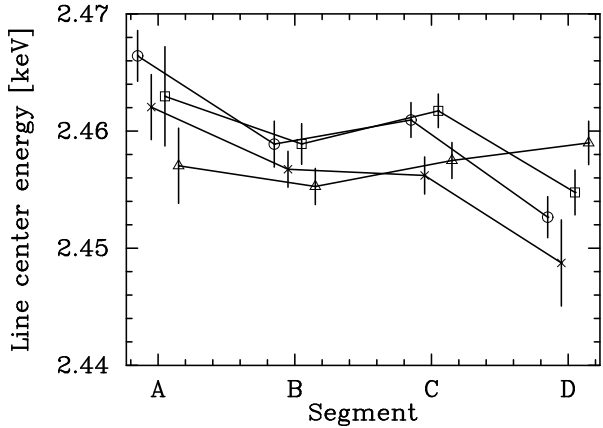} 
     \caption{The center energy of S\emissiontype{XV}~K$\alpha$ after the CTI 
correction in the 4 segments of each XIS. The symbols are the same as figure \ref {fig:4xis-cti0Sheka}.}
     \label{fig:4xis-actxSheka}
   \end{center}
 \end{figure}

\begin{figure}[!ht]
   \begin{center}
     \FigureFile(70mm,50mm){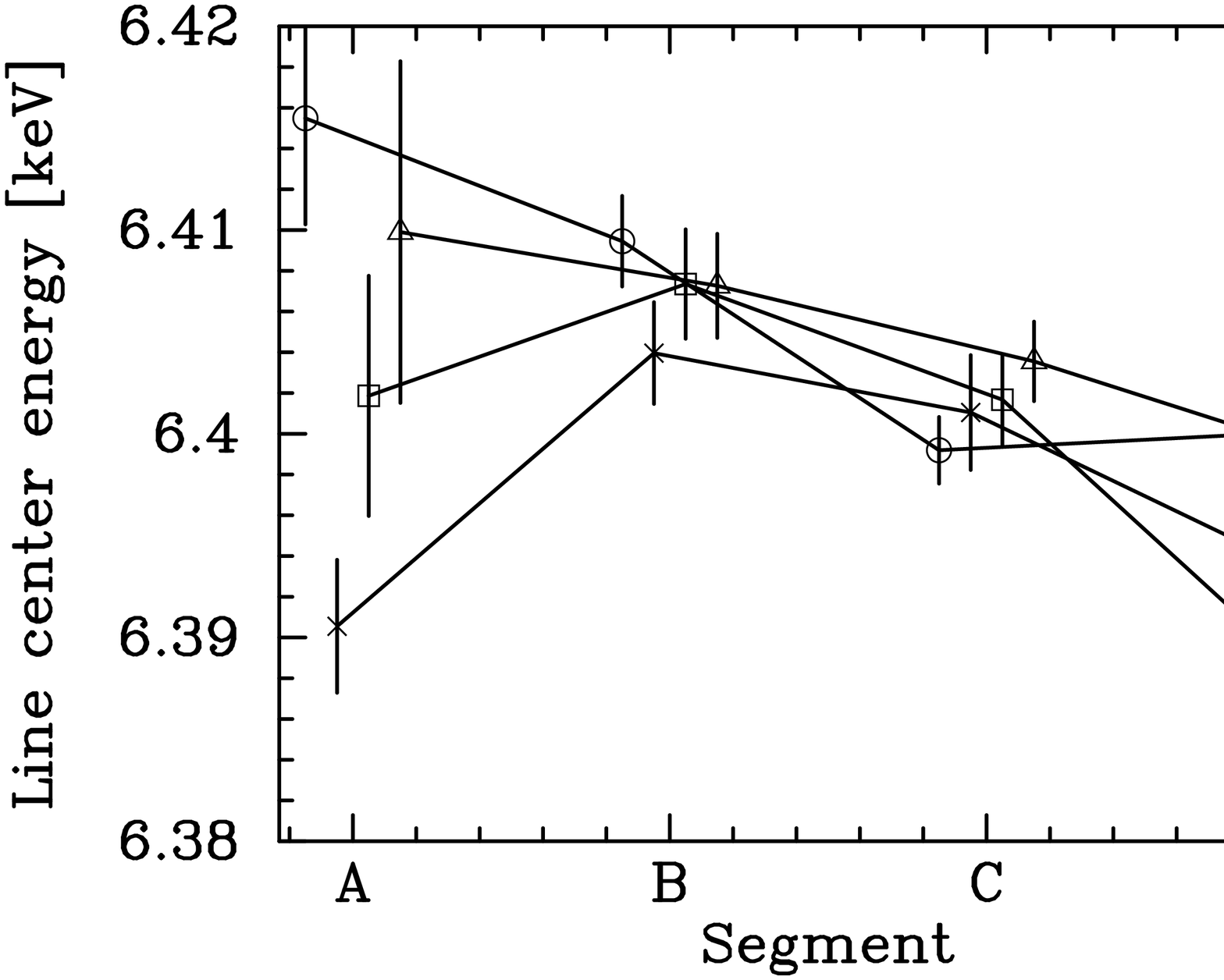} 
     \caption{Same as figure \ref{fig:4xis-actxSheka}, but for the of Fe\emissiontype{I}~K$\alpha$ line.}
     \label{fig:4xis-actxFeneuka}
   \end{center}
 \end{figure}

\begin{figure}[!ht]
   \begin{center}
     \FigureFile(70mm,50mm){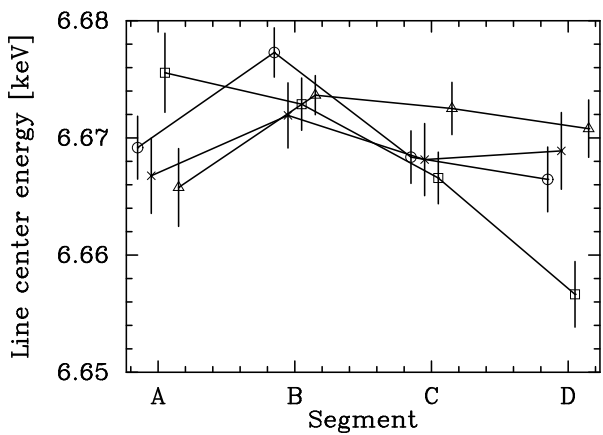} 
     \caption{ Same as figure \ref{fig:4xis-actxSheka}, but for the of Fe\emissiontype{XXV}~K$\alpha$ line.}
     \label{fig:4xis-actxFeheka}
   \end{center}
 \end{figure}

We then  plot the CTI-corrected data for the 4 segments (segment A, B, C and D) of each XIS, and the results are shown in figures \ref{fig:4xis-actxSheka}, \ref{fig:4xis-actxFeneuka} and \ref{fig:4xis-actxFeheka}.
 

\newpage

\begin{figure}[!ht]
   \begin{center}
     \FigureFile(70mm,50mm){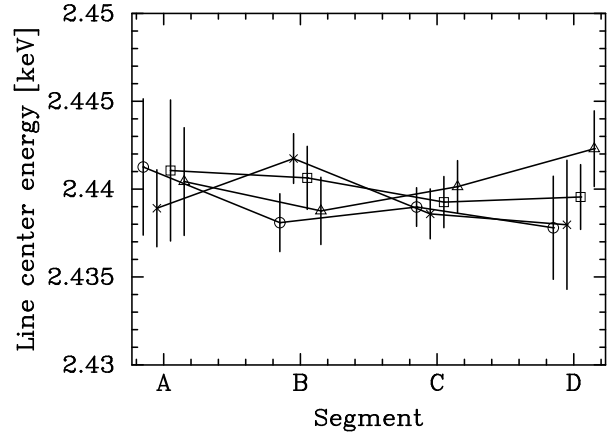} 
     \caption{The same as figure \ref{fig:4xis-actxSheka}, but after the gain correction among the each segment. The symbols are the same as figure \ref{fig:4xis-cti0Sheka}.}
     \label{fig:4xis-gainSheka}
   \end{center}
 \end{figure}

\begin{figure}[!ht]
   \begin{center}
     \FigureFile(70mm,50mm){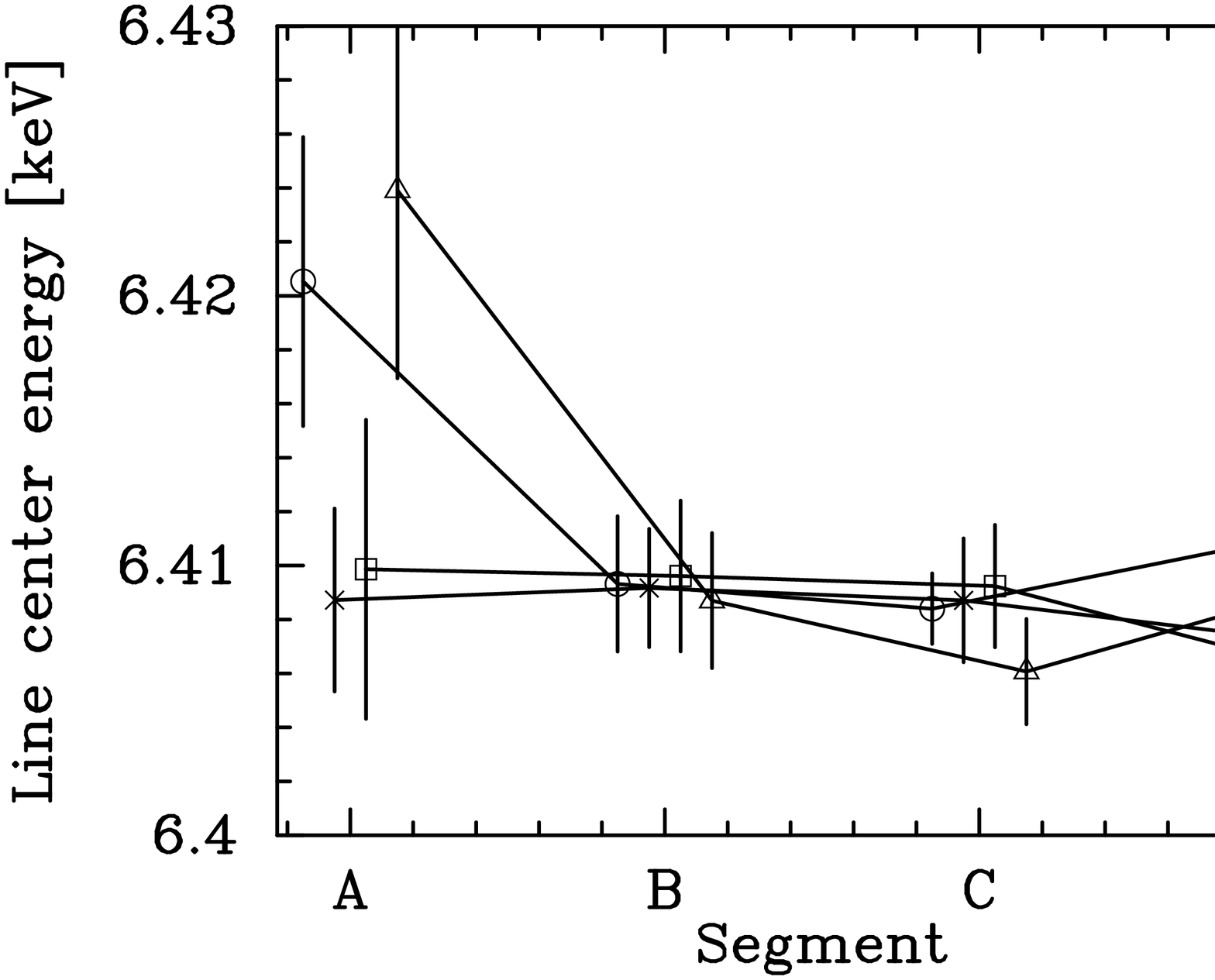} 
     \caption{Same as figure \ref{fig:4xis-gainSheka}, but for the Fe\emissiontype{I}~K$\alpha$ line.}
     \label{4xis-gainFeneuka}
   \end{center}
 \end{figure}

\begin{figure}[!ht]
   \begin{center}
     \FigureFile(70mm,50mm){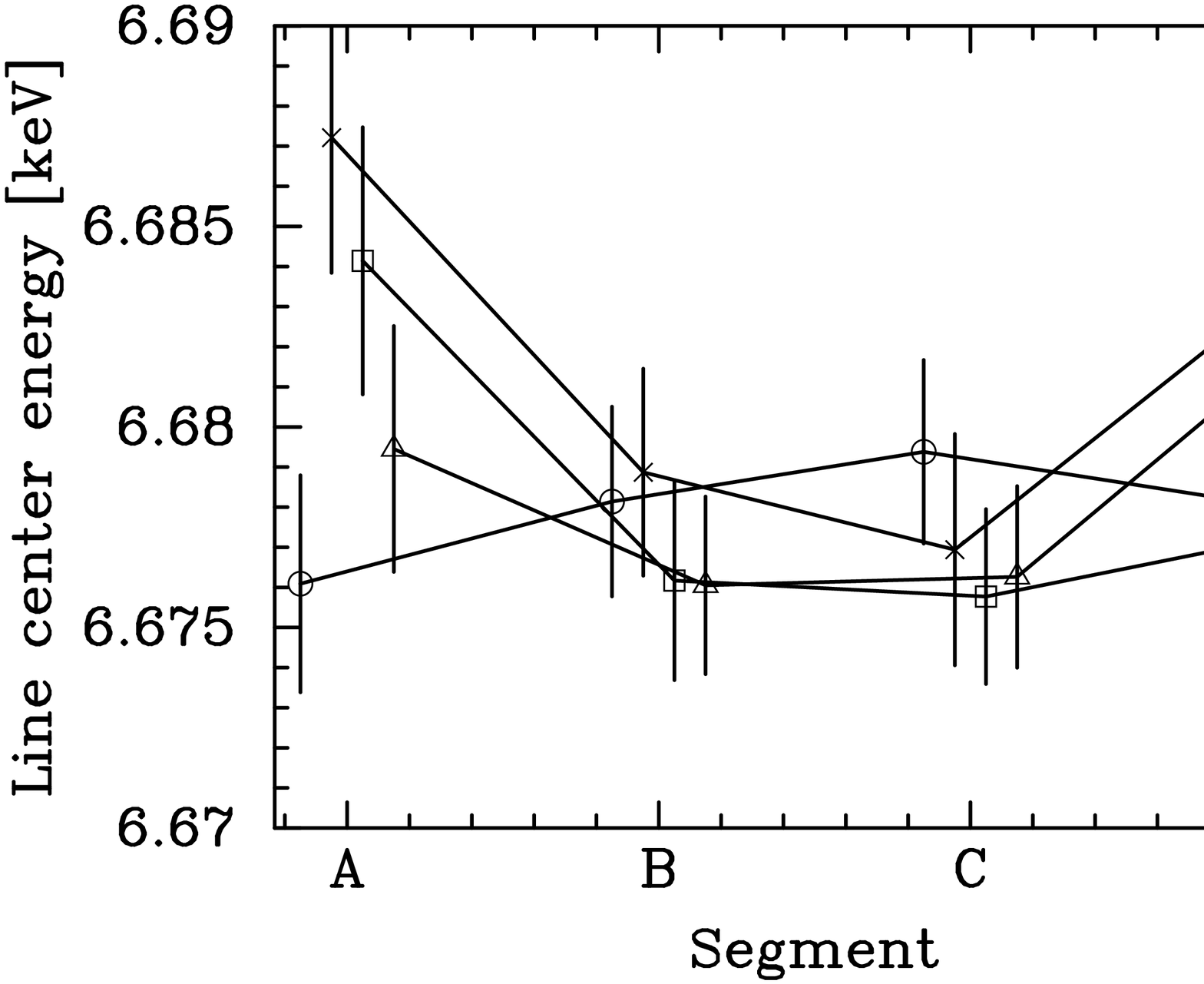} 
     \caption{Same as figure \ref{fig:4xis-gainSheka}, but for the Fe\emissiontype{XXV}~K$\alpha$ line.}
     \label{fig:4xis-gainFeheka}
   \end{center}
 \end{figure}

We find significant gain differences from segment to segment in each XIS
and from XIS to XIS in the segment-averaged energy. The absolute gains for the segments including the calibration sources are fine-tuned to the energy of the calibration lines (K$\alpha$ and K$\beta$ lines of Mn\emissiontype{I})  The gains of the other segments are adjusted to match the center energies of the 3 K-shell lines (S\emissiontype{XV}, Fe\emissiontype{I} and Fe\emissiontype{XXV}) to those of the fine-tuned energies  of the  calibration segments.  
The results are given in figures \ref{fig:4xis-gainSheka}, \ref{4xis-gainFeneuka}
 and \ref{fig:4xis-gainFeheka}.
Note that the vertical axis scales are finer than those of figures \ref{fig:4xis-cti0Sheka}--\ref{fig:4xis-actxFeheka}.  We see the energy variations from XIS to XIS  and/or from segment to segment become smaller than those of 
the non-correction in both the S and Fe K$\alpha$-lines.   

\end{document}